\documentclass[onecolumn,aps,prc,floatfix,showpacs,showkeys,superscriptaddress]{revtex4-1}
\usepackage{amsfonts}
\usepackage{amssymb}
\usepackage{amsmath}
\usepackage{amssymb}
\usepackage{graphicx}
\usepackage{dcolumn}
\usepackage{float}

\usepackage{hyperref}
\usepackage{appendix}
\usepackage{enumitem}
\begin{document}

\title{ Derivation of the Strutinsky method from the least squares principle}
\author{B. Mohammed-Azizi}
\email{aziziyoucef@gmail.com}

\affiliation{Laboratoire de Physique des Particules et Physique Statistique, Ecole Normale Superieure de Kouba, BP 92, Vieux-Kouba, 16050 Algiers, Algeria}
\affiliation{University of Bechar, Bechar, Algeria}
\date{\today }
\begin{abstract}
The main purpose of this paper is to rigorously establish the Strutinsky
method from the least squares principle. Thus, it is the mathematical
basis of this method (aspect often neglected) which is revisited in
an extensive way. Some formulas previously given without demonstration
or in a simplified way are set out here with all the details. In this
respect, the most important mathematical properties of the averaging
functions are also established in this paper. When some conditions
are met, it turns out that Strutinsky's method is nothing more than
a polynomial moving average of the semi-classical level density.
\end{abstract}

\keywords{Nuclear theory, Nuclear structure, Strutinsky averaging method, Binding
energy }

\pacs{21.60.-n, 21.10.Ma, 21.10.Dr, 02.60.-x}
\maketitle

\section{Introduction}

In nuclear structure theory, the macroscopic-microscopic model is
closely related to Strutinsky's discovery of the so called shell correction
method \cite{strut 1967,strut 1968}. This model constitutes a powerful
analytical tool in the study of fission barriers, equilibrium forms
of nuclei and nuclear masses \cite{brack 1972,nilsson 1969,bolsterli 1972,nix 1972}
as well as in other domains \cite{reimann 1993,poenaru 2008,baojing 2007}.
It continues to be applied until today \cite{poenaru 2015,adamian 2018,fan 2017,zhiming 2019},
especially in extensive or systematic studies. In fact, it is to be
noted that in the literature the macroscopic-microscopic method and
shell correction mean the same thing. However, it should also be remembered
that attempts to quantify shell correction took place before Strutinsky's
discovery. In this respect, we can mention the Myers-Swiatecki version
of the shell correction which was roughly defined from the notion
of the energy level bunching \cite{myers 1966}.

The main strength of the macroscopic-microscopique method is that
it combines a strong interaction model (the liquid drop model) with
a weak interaction model (the shell model). This is due the fact that
the advantages of the two models add up while their weaknesses offset
each other. This association simply consists in taking the energy
of the liquid drop model and in adding to it an energy correction
due to the shell effects. The foundations of such an association of
two very different models have been justified microscopically in Ref.
\cite{bunatian 1972,brack 1981} using the Hartree-Fock method and
the Fermi liquid theories.

The key element of the macroscopic-macroscopic method lies in the
original idea of Strutinsky to extract a shell correction from the
shell model. This mathematical procedure is not uncritical. From a
practical point of view, one of the most important question is that
of the plateau issue. This problem has been discussed in several references
\cite{wing 16,ramamurthy 17,niu 18,salamon 19,pomorski20,azizi 21}.

The second main criticism of Strutinsky's method lies in the fact
that the function which averages (and smoothes) the quantum density
of states is built only from an ansatz, i.e. from an unfounded basis.
It is therefore given in a purely intuitive way. Originally, Strutinsky's
idea was to replace Dirac's delta functions of the quantum density
of states by Gaussians so as to get a continuous and regular level
density (provided that the Gaussians are sufficiently wide). The curvature
correction of the method was subsequently added because it was realized
that the method was \char`\"{}missing something\char`\"{} \cite{strut 1967}.
But this way of doing is more like a recipe than a rigorous scientific
analysis. Later, the mathematical formalism of the Strutinky procedure
evolved since the definition of smoothing will be based on a convolution
product (see section\ref{section 4}). But even this convolution product
is also given in an ad hoc way.

To remedy to this problem the present paper first proposes to demonstrate
that Strutinsky level density can be precisely defined as a local
polynomial smoothing of the quantum level density in the least-squares
sense . In fact, this definition alone is not sufficient because the
smoothing is effectively done only in a fixed window (region) of the
variable. To generalize the smoothing of the level density $g(\epsilon)$
at any point $\epsilon$, one has to perform a moving average (the
interval of the averaging is displaced with $\epsilon$). When this
is is done, we recover rigorously the Strutinsky method. A moving
(or rolling) average is usually a numerical technique (for discrete
data) used in mathematics, statistics or in electronics, but here
it is used in analytical way. The Strutinsky level density is thus
rigorously established on the basis of the principle of the least
squares method.\\
 Moreover, it has been proved that within the so-called asymptotic
limit, the Strutinsky density is an approximation of the semi-classical
method \cite{mazizi 2006}. By comparing this result with the result
of the present paper we can conclude that in this limit the density
deduced by Strutinsky's method is a moving polynomial average of the
semi-classical density.

This paper is presented as follows: In section \ref{section 2} it
is shown how to split the quantum level density into a smooth part
and a rapid variations with the inherent hypotheses. Then the section
\ref{section 3} establishes the local polynomial approximation of
the quantum density on the basis of the least square principle (first
part of the proof ) and shows how to generalize this averaging to
any point $\epsilon$ of the level density $g(\epsilon)$ by making
a polynomial moving average (second part of the proof). The usual
integral form of the average density of levels (which is given as
a definition in most papers) is established in section \ref{section 4}
from the result of the previous section. Section \ref{section 5}
is devoted to the main properties of the averaging function. Section
\ref{section 6} interprets the Strutinsky method as a polynomial
approach to semi-classical density in the least-squares sense. A simple
example illustrating the error made by Strutinsky's method is given
in section \ref{section 7}. The main points to remember are given
in the conclusion (section \ref{section 8}).

\section{Smooth part of the quantum level density\label{section 2}}

In quantum physics, the density of states $g(\epsilon)$ is defined
as the number of states per energy unit. These states are deduced
from the solution of the time-independent Schrodinger equation. For
a finite well, there is a finite number of bound states within the
well and a continuum above. The level density is then defined as:
\begin{equation}
g(\epsilon)=\sum_{i=0}^{\nu}\delta(\epsilon-\epsilon_{i})+g_{cont}(\epsilon)\label{eone}
\end{equation}
For an infinite potential there are only discrete (bound) states:
\begin{equation}
g(\epsilon)=\sum_{i=0}^{\infty}\delta(\epsilon-\epsilon_{i})\label{etwo}
\end{equation}
In this respect, it is useful to make the following remarks. From
a practical point of view, the continuum is rather difficult to solve.
In most of cases, the resolution of the Schrodinger equation is done
through a discrete harmonic oscillator basis and all the spectrum
is obtained under a discrete form. In, principle in this way, only
bound states are to be considered and the continuum, obtained thus
under a discrete form, is not valid and must be solved in other ways.
This makes things difficult . Fortunately, there is a method \cite{vertse 2000-1}
which avoid that, solving the level density by using also unbound
states in their discrete form. In this way Eq. (\ref{eone}) can reduces
to the more simple form given by Eq. (\ref{etwo}). Here, in the rare
cases where formula (\ref{etwo}) is used, it will be assumed that
the level density of states has been determined by the method of reference
\cite{vertse 2000-1}. \\
 Regardless on how this function is solved, in the following, we only
will assume that the function $g(\epsilon)$ is known (i.e. already
solved), and that, it is not necessary to specify its explicit form.\\
 It is also assumed that this density of states results from the superposition
of a smooth monotonous function $g_{0}(\epsilon)$ and a fluctuating
or oscillating part $\delta g(\epsilon)$. This can be written as:
\begin{equation}
g(\epsilon)=g_{0}(\epsilon)+\delta g(\epsilon)\label{eq:one}
\end{equation}
It is known that $g_{0}(\epsilon)$ is monotonously increasing function
for infinite wells whereas for finite wells it is monotonously increasing
and then monotonously decreasing \cite{vertse 2000-1}. It will be
assumed that, $g(\epsilon)$ and $g_{0}(\epsilon)$ have the same
asymptotic behavior for the increasing part. This is conditioned by
the so called asymptotic limit \cite{azizi 21,mazizi 2006} : 
\begin{equation}
g(\epsilon)\sim g_{0}(\epsilon)\ \ as\ \ \epsilon\gg\hbar\omega\label{eq:asymp}
\end{equation}
where $\hbar\omega$ represents one quantum of the energy of the system.
In fact within this limit the density of the energy levels is so large
that it can be considered as continuous. This justifies the application
of semi classical methods.\\
 The aim of the proposed method is only to determine the smooth part
$g_{0}(\epsilon)$ which does not contain oscillations and which can
be considered as close to semi classical level density $g_{sc}(\lambda)$.
As we will see, the mathematical process we are going to use, leads
to the Strutinsky method.

\section{Derivation of the Strutinsky procedure as a polynomial approximation
from a local least squares principle \label{section 3}}

The proof will be established in two successive steps.

\subsection{Step 1: Local least squares polynomial fit}

In a first step, we want to average or to smooth $g(\epsilon)$ by
a polynomial $Q_{M}(\epsilon)$ of degree $M$, in the vicinity of
some point $\epsilon=\lambda$ over a range of values weighted by
a Gaussian function of effective width $\gamma$. The least squares
principle will be applied by minimizing the following integral which
represents a sum of squared errors: 
\begin{equation}
I(M,\gamma,\lambda)=\intop_{-\infty}^{+\infty}\left\{ g(\epsilon)-Q_{M}(\epsilon)\right\} ^{2}e^{-(\frac{\epsilon-\lambda}{\gamma})^{^{2}}}d\epsilon\label{eq:grand I}
\end{equation}
Thus polynomial $Q_{M}(\epsilon)$ averages the function $g(\epsilon)$
locally over an effective range defined by the Gaussian parameter$\gamma$
around $\lambda$. By construction, the coefficients of that polynomial
depend a priori on the degree $M$ (arbitrarily fixed) and because
of the Gaussian, they also depend on $\gamma$ and $\lambda$. Furthermore,
due to the Gaussian weight, the most interesting form is to write
this polynomial as a combination of Hermite's polynomials of degree
$m$ (up to $M$):

\begin{equation}
Q_{M,\gamma,\lambda}(\epsilon)=\sum_{_{m=0}}^{^{M}}d_{m}(M,\gamma,\lambda)H_{m}\left(\frac{\epsilon-\lambda}{\gamma}\right)\label{eq:peem3}
\end{equation}
The coefficients ensuring the least squares principle will be resolved
by minimization: 
\begin{equation}
\frac{\partial I}{\partial d_{k}}=0\label{eq:deriv}
\end{equation}
By Using the orthogonality property of Hermite polynomials, we finds
the following coefficients: 
\begin{equation}
d_{k}(\gamma,\lambda)=\frac{1}{2^{k}k!\sqrt{\pi}\gamma}\intop_{-\infty}^{+\infty}g(\epsilon)H_{k}\left(\frac{\epsilon-\lambda}{\gamma}\right)e^{-\left(\frac{\epsilon-\lambda}{\gamma}\right)^{2}}d\epsilon\label{eq:coeff}
\end{equation}
As it can be easily seen, index $M$ does not appear in the RHS of
Eq. (\ref{eq:coeff}), this is the reason why it has been suppressed
from coefficient $d_{k}(\gamma,\lambda)$\\
 Note: In the relationship (\ref{eq:grand I}), a Gaussian weight
was used. In fact any weight function tending towards the Dirac delta
function can be used. For example, it is perfectly possible to replace
the Gaussian weight function with a Lorentz function in order to apply
the least squares principle in the same way. In that respect, references
\cite{brack 1973,magnus 2008} can be consulted. In the present work,
for convenience and simplicity, a Gaussian weight has been chosen.
This is mainly due to the ease of use of Hermite polynomials (which
are associated to the Gaussian weight).

\subsection{Step 2: Definition of the Strutinsky level density as a least squares
moving average}

In fact when we want to calculate a local average of $g(\epsilon)$
at a given point $\epsilon$ we have to make the average with the
values of $g(\epsilon)$ that precede and that follow the point $\epsilon$
within a given window. In our case the window is defined by the Gaussian
weight which favours the values of $g(\epsilon)$ that are close to
$\lambda$. Thus, due to the Gaussian weight, the averaging of $g(\epsilon)$
is inflenced mainly by the values which are near $\lambda$ (fixed
in the previous step) and not by the ones that are near $\epsilon$.
In order to get a true local average of $g(\epsilon)$ in which the
points that precede and that follow the point $\epsilon$ play the
main role, it is necessary to make the center of the bell curve (Gaussian)
coincide with the point $\epsilon$ in which the average density of
$g(\epsilon)$ is made. To evaluate the averaging of $g(\epsilon)$
everywhere, the bell curve (Gaussian window) must then be moved continuously
with the point $\epsilon$. Averaging that function means that, at
each point $\epsilon$, the value of $g(\epsilon)$ must replaced
by the weighted average of its neighboring values (near $\epsilon$).
It turns out that this process is nothing more than a moving average
of function $g(\epsilon)$ or more precisely a polynomial moving average.
In practice, this simply amounts to replacing straightforwardly $\epsilon$
by $\lambda$ (or conversely $\lambda$ by $\epsilon$) in the polynomial
obtained by means of the least squares principle (\ref{eq:peem3}):
\begin{equation}
\left.Q_{M,\gamma,\lambda}(\epsilon)\right|_{\epsilon=\lambda}=\sum_{_{k=0}}^{^{M}}d_{k}(\gamma,\lambda)H_{k}(0)\label{eq:dens}
\end{equation}
The choice made here amounts to taking $\lambda$ (instead of $\epsilon$)
as variable of $g$. In this case, the least square approximation
of $g(\lambda)$ reads:

\[
g(\lambda)\approx Q_{M,\gamma,\lambda}(\lambda)
\]
In the following, due to the redundancy of $\lambda$ the Strutinsky
density will be simply represented by $Q_{M,\gamma}(\lambda)$. It
must be noted that the $Q_{M,\gamma,\lambda}(\lambda)$ polynomial
approximation obtained in this way, is local, i.e. it is valid only
for the point $\lambda$. For another point $\lambda'$, the local
polynomial approximation $Q_{M,\gamma,\lambda'}(\lambda')$ remains
valid but the two polynomials are not identical . In other words the
coefficients of this polynomial change continuously with the point
$\lambda$. This is essentially due the moving average introduced
in the second part of the proof which makes the Strutinsky method
somewhat ``subtle'' (see also note in section \ref{section 5}).

In order to check that Eq. (\ref{eq:dens}) give the usual well known
result let us replace in Eq. (\ref{eq:coeff}) $g(\epsilon)$ by its
value from Eq. (\ref{etwo}), using the property of Dirac distribution,
we find:

\[
d_{k}(\gamma,\lambda)=\frac{1}{2^{k}k!}\,\sum_{i=0}^{\infty}H_{k}\left(\frac{\epsilon_{i}-\lambda}{\gamma}\right)\frac{1}{\gamma\sqrt{\pi}}e^{-\left(\frac{\epsilon_{i}-\lambda}{\gamma}\right)^{2}}
\]
Inserting this result in Eq. (\ref{eq:dens}), we obtain:

\begin{equation}
Q_{M,\gamma}(\lambda)=\sum_{i=0}^{\infty}\left\{ \sum_{k=0}^{M}B_{k}H_{k}(x_{i})\right\} \frac{1}{\gamma\sqrt{\pi}}e^{-x_{i}^{2}}\label{strut}
\end{equation}
with $B_{k}=H_{k}(0)/2^{k}k!$, and $x_{i}=(\epsilon_{i}-\lambda)/\gamma$.
$H_{k}(0)$ is given in appendix \ref{appx b}. We recover the usual
well-known ``classical'' result of Strutinsky, i.e. a Gaussian multiplied
by the polynomial curvature correction \cite{nix 1972}. Thus Strutinsky
density $Q_{M,\gamma}(\lambda)$ results from a polynomial moving
average\\
 The moving average defined above does not ensure that $Q_{M,\gamma}(\lambda)$
be a regular function. Thus, to smooth that quantity it is necessary
to enlarge the width of the Gaussian in such way that it must be at
least of the order of the average shell-spacing $\hbar\omega$ \cite{strut 1967,brack 1972,nilsson 1969,bolsterli 1972}

\begin{equation}
\gamma\succsim\hbar\omega\label{smoothing-1}
\end{equation}
This is the well known smoothing condition of the Strutinsky procedure.

\section{Derivation of the integral form of the moving average. Averaging
functions\label{section 4}}

Coefficient $d_{k}$ is given by Eq. (\ref{eq:coeff}). Replacing
its expression in formula (\ref{eq:dens}), making $x=\left(\epsilon-\lambda\right)/\gamma$
and inverting sum and integral signs, we get an equivalent form of
the smoothing average (i.e., the Strutinsky density) cited above (Eq.
(\ref{eq:dens})) as follows: 
\begin{equation}
Q_{M,\gamma}(\lambda)=\intop_{-\infty}^{+\infty}g(\lambda+\gamma x)F_{M}(x)dx\label{eq:ggam-1}
\end{equation}
where the averaging (or smoothing) function is defined by:\\

\begin{equation}
F_{M}(x)=\left\{ \sum_{m=0}^{^{M}}B_{m}H_{m}(x)\right\} \frac{e^{-x^{2}}}{\sqrt{\pi}}\label{eq:fm}
\end{equation}
with $B_{m}=H_{m}(0)/2^{m}m!$\\
 There is a second form for the averaging function (see appendix \ref{appx c}):
\begin{equation}
F_{M}(x)=\left\{ B_{M}\frac{H_{M+1}(x)}{2x}\right\} \frac{e^{-x^{2}}}{\sqrt{\pi}}\label{eq:fm2-1}
\end{equation}
Moreover, an other integral form of Eq. (\ref{eq:ggam-1}) can be
obtained by making again $x=(\epsilon-\lambda)/\gamma$ in that result:
\begin{equation}
Q_{M,\gamma}(\lambda)=\intop_{-\infty}^{+\infty}g(\epsilon)\frac{1}{\gamma}F_{M}(\frac{\epsilon-\lambda}{\gamma})d\epsilon\label{eq:gstrut}
\end{equation}
Here, we find the usual form of the convolution product of the Strutinsky
method. In this respect, in mathematics, it is well known that there
is a close relationship between the convolution product and the moving
average. \\
 In the integral form of the averaging of Eq. (\ref{eq:gstrut}) we
can replace the quantum level density by its definition from Eq. (\ref{etwo})
Applying property of Dirac distribution one gets:

\begin{equation}
Q_{M,\gamma}(\lambda)=\sum_{i=0}^{\infty}\frac{1}{\gamma}F_{M}(\frac{\epsilon_{i}-\lambda}{\gamma})\label{gpi}
\end{equation}
It must be noted that Eq. (\ref{gpi}) and Eq. (\ref{strut}) are
identical. This result is achieved in two different ways and constitutes
a check of this formula. We also have to add the same smoothing condition
(\ref{smoothing-1}) as seen before.

\section{Properties and width of the averaging function\label{section 5}}

The basic properties of the averaging functions $F_{M}(x)$ are given
in appendix \ref{appx 1}. Most of them are necessary for understanding
or demonstrations The non obvious properties are demonstrated in Appendix
\ref{appx d}.\\
 The most fundamental property is the following integral transform
(giving the the Strutinsky density) already seen in Eq. (\ref{eq:gstrut})
in which $g(\epsilon)$ represents here any function (not necessarily
a level density):

\begin{equation}
\intop_{-\infty}^{+\infty}g(\epsilon)\frac{1}{\gamma}F_{M}(\frac{\epsilon-\lambda}{\gamma})d\epsilon=g(\lambda)+\sum_{k=M+2}^{\infty}D_{kM}\gamma^{k}\frac{d^{k}g(\lambda)}{d\lambda^{k}}\label{transf}
\end{equation}
In which:

\[
D_{kM}=\left(\frac{(-1)^{M/2}}{(M/2)!}\right)\left(\frac{1}{\left[(k-M-2)/2\right]!}\right)\left(\frac{1}{2^{k-1}k}\right),\,\,\,k,\,M=even
\]
This formula is obtained by making a Taylor expansion of $g(\lambda+\gamma x)$
in Eq. (\ref{eq:ggam-1}) and using properties number 5, 6 and 9 of
the table given in appendix \ref{appx 1}. From this formula some
consequences can be deduced: 
\begin{itemize}
\item Because of the derivatives in Eq. (\ref{transf}), if $g(\epsilon)$
is a polynomial of degree $M$ (or less), the series vanishes and
the integral transform of Eq. (\ref{transf}) gives back to the same
polynomial $g(\lambda)$ (in $\lambda$). So if $g(\epsilon)$ is
a polynomial this integral returns the same polynomial (in $\lambda$).
Conversely, if $g(\epsilon)$ is not a polynomial, this transformation
gives back again to $g(\lambda)$ but with a remainder which is of
the order of the first term of the series (for which $k=M+2$). 
\item If $\gamma=0$, the series in the right-hand side of Eq. (\ref{transf})
cancels. This means that integral in Eq. (\ref{transf}) gives back
to the same function. In other words $\frac{1}{\gamma}F_{M}(\frac{\epsilon-\lambda}{\gamma})$
reduces to Dirac function in the limit $\gamma\rightarrow0$. Analogous
property holds if $M\rightarrow\infty$. Indeed, in this case, the
remainder of the Taylor expansion of $g(\epsilon)$ in the integral
of Eq. (\ref{transf}) also cancels involving the same consequence
for the series in Eq. (\ref{transf}) (which is in fact, the integral
of this remainder). So, in that situation, $\frac{1}{\gamma}F_{M}(\frac{\epsilon-\lambda}{\gamma})$
reduces also to Dirac function. 
\item As regard the real width of the averaging functions (see also points
just below), the above properties shows that, the actual width of
the function $F_{M}(x)$ is not due to the sole $\gamma$ parameter
(which characterizes the width of the Gaussian) but is governed by
the both parameters $\gamma$ and $M$. It increases with $\gamma$
and diminishes with $M$. 
\item An other notable property is that the area of this function is always
normalized to the unit. This can be shown by simply making $g(\epsilon)=1$
in Eq. (\ref{transf}). Its main maximum (in $x=0$) increases when
its width decreases and vice versa. Thus: 
\end{itemize}
\begin{equation}
\intop_{-\infty}^{+\infty}\frac{1}{\gamma}F_{M}(\frac{\epsilon-\lambda}{\gamma})d\epsilon=1\label{unit}
\end{equation}

\begin{figure}[H]
\centering \includegraphics[width=0.58\columnwidth,keepaspectratio]{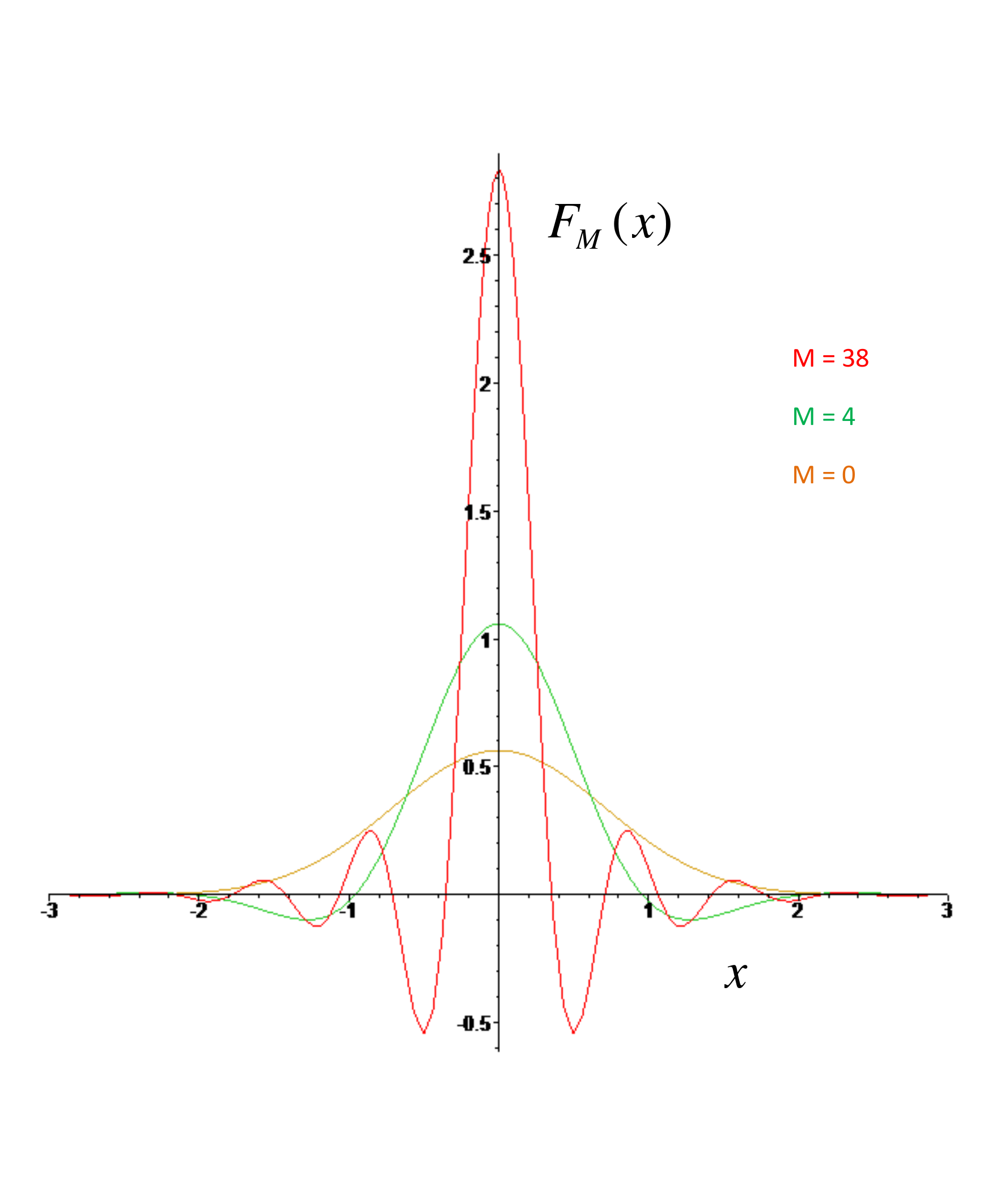}
\caption{ Averaging functions Eq.(\protect (\ref{eq:fm2-1})) for three values
of the degree $M$. It is to be noted that these functions practically
cancel as soon as $\left|\right|\gtrsim2$}
\label{figsmooth} 
\end{figure}

On the other hand, without going into further details, one also can
say that: 
\begin{itemize}
\item Due to the Hermite polynomial in the second representation given by
Eq. (\ref{eq:fm2-1}), the averaging function has $M$ symmetrical
roots with respect to the $x=0$ axis, $(M/2)$ positive and $(M/2)$
negative with $M$ even. It has a main maximum in $x=0$ and successive
oscillations with an amplitude that decreases rapidly (see Fig. \ref{figsmooth})). 
\item The main contribution of this area comes from the central part because
these oscillations decrease rapidly in amplitude and also because
they have successive areas that are of opposite signs and thereby
compensate each other. In terms of area, the curve of $F_{M}(x)$
is practically confined between $-1$ and $+1$. The following property
is all the more true as the quantity $a$ is large compared to the
unit. 
\begin{equation}
\intop_{-a}^{+a}F_{M}(x)dx\approx1\ \ \ if\ \ \ a\gg1\,\,\,\,\forall M\label{xgrand}
\end{equation}
In effect, it is to be noted that even for $a=1$, the precision is
already satisfactory. For example, taking $M=30$ and $a=1$ the value
of the integral will be about 0.9914. For $M=2$ and $a=1$, one obtains
1.050. Because the oscillations change sign, the values of the area
are given either by excess or by default.\\
 Curve $(1/\gamma)F_{M}(x/\gamma)$ has an analog property as the
one given by Eq. (\ref{xgrand}), i.e. are practically confined between
$-\gamma$ and $+\gamma$ . In other words, it has a ``quasi-width''
of length about $2\gamma$. 
\end{itemize}


\section{The relative remainder of the averaging in the asymptotic limit\label{section 6}}

Because the right-hand side of Eq. (\ref{eq:gstrut}) is equal to
the Strutinsky density $Q_{M,\gamma}(\lambda)$, we can deduce the
link between the Strutinsky level density $Q_{M,\gamma}(\lambda)$
and the quantum density $g(\lambda)$. Indeed, rewriting Eq. (\ref{transf})
as follows: we find that:

\begin{equation}
Q_{M,\gamma}(\lambda)=g(\lambda)\left(1+R_{M+2,\gamma}(\lambda)\right)\label{eq:pg}
\end{equation}
where the relative remainder is the sum of as a Taylor remainder divided
by $g(\lambda)$. This relative remainder is of the order of the first
element of this sum, i.e:

\begin{equation}
R_{M+2,\gamma}(\lambda)\approx\frac{(-1)^{M/2}}{(M/2)!2^{M+1}(M+2)}\left(\frac{g^{(M+2)}(\lambda)}{g(\lambda)}\right)\gamma^{M+2}\label{eq:remainder}
\end{equation}
in which $g^{(M+2)}(x)$ represents the $(M+2)th$ derivative ($M$
even). Because actually this formula comes from the Taylor expansion
of $g(\lambda+\gamma x)$, the Taylor remainder will be small only
if condition $\lambda\gg\gamma\left|x\right|$ is well satified. Due
to property \ref{xgrand}, this condition turns to be:

\begin{equation}
\lambda\gg\gamma\label{asymptotic}
\end{equation}
which in fact is equivalent to the asymptotic limit defined in section
\ref{section 2} (where variable $\epsilon$ has simply been replaced
by $\lambda$): 
\begin{equation}
\lambda\gg\hbar\omega\label{same}
\end{equation}
because $\gamma$ and $\hbar\omega$ are of the same order. To show
that, it is sufficient to note that the compatibility between condition
\ref{smoothing-1} and asymptotic limit \ref{asymptotic} can be summarized
in only one formula:

\begin{equation}
\lambda\gg\gamma\gtrsim\hbar\omega\label{eq:doco}
\end{equation}
Classically, $\lambda$ and $\hbar\omega$ are connected to physical
conditions, thus the only free parameter, i.e. $\gamma$ should be
neither too large nor too small. The optimal choice is then necessarily
about $\gamma\approx\hbar\omega$, in this case Eq. (\ref{asymptotic})
and condition (\ref{same}) are thus equivalent.

Consequently, if conditions of Eq. (\ref{eq:doco}) are fulfilled,
$g(\lambda)$ will be smooth, like $g_{0}(\lambda)$ defined in section
\ref{section 2} and will be close to the semi classical density (see
Ref. \cite{mazizi 2006}). In addition, under these conditions, the
remainder $R_{M+2}(\lambda)$ will be small. In this case, the Strutinsky
level density can be considered as a local polynomial approximation
to the semi classical density. This is why the Strutinsky procedure
gives the exact result for $M\geq2$ in the case of the three dimensional
harmonic oscillator where the semiclassical level density is rigorously
a polynomial of degre two (parabola) \cite{brack 1973}.

\section{Illustrative example of the accuracy of the Strutinsky Method. Particle
in a cubic box with reflecting walls \label{section 7}}

For the cubic box it is possible to estimate the relative error of
the Strutinsky method with respect to semi classical result. For this
example, the eigenvalues are: 
\begin{equation}
\epsilon_{n_{x}n_{y}n_{z}}=(n_{x}^{2}+n_{y}^{2}+n_{z}^{2})E_{0},\,\,\,\,\,\,\,\,\,E_{0}=\frac{\pi^{2}\hbar^{2}}{2ma^{2}}
\end{equation}
herein $n_{x}\ or\ n_{y}\ or\ n_{z}=1,2,.....\infty$ are quantum
numbers (the value $0$ is forbidden because the wave function cancels
in this case), $\hbar$ is the Planck constant and $m$ is the mass
of the particle and $a$ the side of the box.\\
 A Truncated asymptotic series can be obtained from Eq. (\ref{gpi})
by using the Euler MacLaurin formula. (without derivatives and highest
terms). The calculations are somewhat lengthy but simple and include
expansions of binomial series that must be truncated after using property
9 of the table in appendix \ref{appx 1}. For example, taking $M=8$
and $M=30$, after some algebra and some numerical evaluations, the
result of the Strutinsky approximation in the case of the cubic box
for $M=8$ and $M=30$ are :\\
 $Q_{8,\gamma}(\lambda)\approx\frac{\pi\lambda^{1/2}}{4E_{0}^{3/2}}\left(1+10^{-6}\left(\frac{\gamma}{\lambda}\right)^{10}\right)-\frac{3\pi}{8E_{0}}+$$\frac{3}{8E_{0}^{1/2}\lambda^{1/2}}\left(1+10^{-5}\left(\frac{\gamma}{\lambda}\right)^{10}\right)$\\
 $Q_{30,\gamma}(\lambda)\approx\frac{\pi\lambda^{1/2}}{4E_{0}^{3/2}}\left(1+10^{-9}\left(\frac{\gamma'}{\lambda}\right)^{32}\right)-\frac{3\pi}{8E_{0}}+$$\frac{3}{8E_{0}^{1/2}\lambda^{1/2}}\left(1+10^{-7}\left(\frac{\gamma'}{\lambda}\right)^{32}\right)$\\
 For the cubic box, the semi classical result (exact solution) is
well known \cite{badhuri 1971}:\\
 $g_{sc}(\lambda)=\frac{\pi\lambda^{1/2}}{4E_{0}^{3/2}}-\frac{3\pi}{8E_{0}}+\frac{3}{8E_{0}^{1/2}\lambda^{1/2}}$

Thus, the relative error is of the of the order of $10^{-6}\left(\gamma/\lambda\right)^{10}$
and $10^{-9}\left(\gamma'/\lambda\right)^{32}$ for the cited examples.
For realistic cases, the error is not so small, but it is still very
acceptable. In this respect, it can be said that if Strutinsky method
is applied correctly, it gives good results and thus can be considered
as an alternative to semi-classical methods.

According to the remark made in the third point of section \ref{section 5},
in these expressions (for $M=8$ and $M=30$) the value of the smoothing
parameter is not the same ($\gamma'\neq\gamma$) because this parameter
increases with the degree $M$.

\section{Conclusion \label{section 8}}

The work made in this paper proves that the smoothing method early
discovered by Strutinsky and used as a simple prescription can in
fact be founded by imposing the least squares principle in the smoothing
procedure. More specifically, the smoothing of the quantum density
must be made by a local polynomial approximation (with a fixed degree)
by minimizing the squared error. However, this definition is not sufficient
and leads to a ``static average”. In order, to perform the average
at any point of the level density it is necessary to make a polynomial
moving average. This is the reason why the proof is done in two steps.
We can summarize that by stating that the Strutinsky prescription
can be rigorously established starting from a principle of ``a polynomial
moving average in the least squares sense\char`\"{}. In addition,
this paper demonstrates also how to obtain the integral representation
of that average (known also as a convolution product). The averaging
function in the integrand has been extensively studied. Practically,
all the properties of the averaging function have been given.\\
 Last but not least, it turns out that when the smoothing and asymptotic
limit conditions are fulfilled, the Strutinsky method is nothing more
than a polynomial moving average of the semi classical level density.
Consequently, Strutinsky's method can be a good alternative to semi-classical
methods.

\appendix

\section{Main properties of the averaging (or smoothing) functions\label{appx 1}}

The main properties of the averaging (or smoothing) functions are
given in the following table:

\begin{table}[H]
\begin{centering}
\begin{tabular}{c|ccc}
 & Property  & Condition  & \tabularnewline
\hline 
\hline 
1  & $F_{M}(x)=0$  & $M\ odd$  & due to the def. of $B_{M}$\tabularnewline
\hline 
2  & $F_{M}(-x)=F_{M}(x)$  & $M\ even$  & M even in the following\tabularnewline
\hline 
3  & $\intop_{-\infty}^{+\infty}G(x)F_{M}(x)dx=0$  & $G(x)\ odd\ function$  & for parity reason\tabularnewline
\hline 
4  & $\intop_{-\infty}^{+\infty}F_{M}(x)dx=1$  &  & see section \ref{section 5}\tabularnewline
\hline 
5  & $\intop_{-\infty}^{+\infty}\frac{1}{\gamma}F_{M}(\frac{\epsilon-\lambda}{\gamma})d\epsilon=1$  &  & variable change in 4 \tabularnewline
\hline 
6  & $\intop_{-\infty}^{+\infty}x^{k}F_{M}(x)dx=0$  & $1\leq k\leq M$  & see appendix \ref{appx 2}\tabularnewline
\hline 
7  & $\intop_{-\infty}^{+\infty}x^{k}F_{M}(x-x_{0})dx=x_{0}^{k}$  & $1\leq k\leq M$  & variable change in 6\tabularnewline
\hline 
8  & $\intop_{-\infty}^{+\infty}\epsilon^{k}\frac{1}{\gamma}F_{M}(\frac{\epsilon-\lambda}{\gamma})d\epsilon=\lambda^{k}$  & $1\leq k\leq M$  & variable change in 7\tabularnewline
\hline 
9  & $\intop_{-\infty}^{+\infty}x^{k}F_{M}(x)dx=C_{k,M}$  & $k\geq M+2,\ \ k\ \ even$  & see appendix \ref{appx 2}\tabularnewline
\hline 
 &  & $C_{k,M}=\frac{(-1)^{M/2}(k-1)!}{(M/2)!2^{k-1}\left((k-M-2)/2\right)!}$  & \tabularnewline
\hline 
10  & $\intop_{-\infty}^{+\infty}\left(\epsilon-\lambda\right)^{k}\frac{1}{\gamma}F_{M}(\frac{\epsilon-\lambda}{\gamma})d\epsilon=C_{k,M}\gamma^{k}$  & $k\geq M+2,\ \ k\ \ even$  & variable change in 9\tabularnewline
\end{tabular}
\par\end{centering}
\caption{\label{tab:table-name}Main properties of the averaging functions.}
\end{table}

\section{Hermite polynomials: \label{appx b}}

\subsection{Definition}

$H_{n}(x)=(-1)^{n}e^{x^{2}}\frac{d^{n}}{dx^{n}}e^{-x^{2}}\ \ with\ \ n=0,1,2,....$The
degree of this polynomial is $n$.

\subsection{Orthogonality and norm:}

$\intop_{-\infty}^{+\infty}H_{n}(x)H_{m}(x)e^{-x^{2}}dx=2^{n}n!\sqrt{\pi}\delta_{nm}$where
$\delta_{nm}$is the Kronecker symbol.

\subsection{Parity and special values}

$H_{n}(-x)=(-1)^{n}H_{n}(x)$

$H_{n}(0)=\frac{(-1)^{n}n!}{(n/2)!}\ \ n\ even,\,\,\,\,\,H_{n}(0)=0\ \ n\ odd$

\subsection{Well known property:\label{appx b4}}

$\intop_{-\infty}^{+\infty}x^{k}H_{m}(x)e^{-x^{2}}dx=0\ \ \ for\ k=0,1,2,.......,m-1$\\
 In other words: $H_{m}(x)$ is orthogonal to any polynomial of degree
less than $m$, with $k<m$.

\section{Christopher-Darboux formula applied to averaging functions \label{appx c}}

The Christopher-Darboux formula is:

$\sum_{m=0}^{M}\frac{H_{m}(x)H_{m}(y)}{2^{m}m!}=\frac{1}{2}\frac{H_{M+1}(x)H_{M}(y)-H_{M+1}(y)H_{M}(x)}{2^{M}M!(x-y)}$\\
 Multiplying this equality by $e^{-x^{2}}/\sqrt{\pi}$ , making $y=0$
, we recover $F_{M}(x)$ of Eq. (\ref{eq:fm}). On the other hand,
Hermite polynomial has the property $H_{m}(0)=0$ for odd $m.$ Consequently
the sum contains only terms corresponding to even values $m$ up to
even $M$. The result is:\\
 $F_{M}(x)=\left\{ \sum_{m=0}^{^{M}}B_{m}H_{m}(x)\right\} \frac{e^{-x^{2}}}{\sqrt{\pi}}=\left\{ B_{M}\frac{H_{M+1}(x)}{2x}\right\} \frac{e^{-x^{2}}}{\sqrt{\pi}}$\\
 Where the constant $B_{M}$is defined by: \\
 $B_{P}=\frac{(-1)^{P/2}}{2^{P}(P/2)!}\ \ P\ even,\,\,\,\,\,\,\,\,\,\,B_{P}=0\ \ P\ odd$

\section{Demonstrations of some Formulas: \label{appx d}}

\subsection{Orthogonality between monomials and averaging functions}

This comes essentially from the property given by the property of
appendix (\ref{appx b4}).\\
 $\intop_{-\infty}^{+\infty}x^{k}F_{M}(x)dx=0\ \ \ k=1,2,3,....,M\ \ \ (M=even)$In
effect, using the second form of $F_{M}(x)$ given in Eq. (\ref{eq:fm2-1}),
we can write:\\
 $\intop_{-\infty}^{+\infty}x^{k}F_{M}(x)dx=\intop_{-\infty}^{+\infty}x^{k}\left\{ B_{M}\frac{H_{M+1}(x)}{2x}\right\} \frac{e^{-x^{2}}}{\sqrt{\pi}}dx=(B_{M}/2)\intop_{-\infty}^{+\infty}x^{k-1}H_{M+1}(x)\frac{e^{-x^{2}}}{\sqrt{\pi}}dx=0\ \ if\ \ k-1<M+1$.
\\
 Due to the result of appendix (\ref{appx b4}), the previous inequality
amounts to taking $k<M+2$ or $k=2,4,6...M$.

\subsection{Property of Hermite polynomials \label{appx 2}}

An important property (used in this work) of Hermite polynomials is:\\
 $\intop_{-\infty}^{+\infty}x^{k}H_{m}(x)e^{-x^{2}}dx=\frac{\sqrt{\pi}k!}{2^{k-m}\left(\frac{k-m}{2}\right)!}\ \ \ \ if\ \ \ k\geq m\geq0$T\\
 This property has been proved from combinations of the following
equation \cite{mishra 1991}: \\
 $\intop_{-\infty}^{+\infty}x^{2w+m}H_{m}(x)e^{-x^{2}}dx=\frac{2^{m}\varGamma(\frac{2w+m+1}{2})\varGamma(\frac{2w+m+2}{2})}{\varGamma(w+1)}\ \ \ \ w=0,1,2,....$with
the well known property of the Gamma function\\
 $\varGamma(z)\varGamma(z+1/2)=2^{1-2z}\sqrt{\pi}\varGamma(2z)$ \\
 and $\varGamma(n+1)=n!$

\subsection{Important property of averaging function}

Appendix \ref{appx 2} leads to the following formula:\\
 $\intop_{-\infty}^{+\infty}x^{k}F_{M}(x)dx=C_{k,M}=\frac{(-1)^{M/2}}{(M/2)!}\frac{(k-1)!}{2^{k-1}}\frac{1}{\left((k-M-2)/2\right)!}\ \ \ k=M+2,\ M+4,\ M+6,.........\infty$\\
 Using the second form of $F_{M}(x)$, i.e. Eq. (\ref{eq:fm2-1}),
we obtain:\\
 $\intop_{-\infty}^{+\infty}x^{k}F_{M}(x)dx=\intop_{-\infty}^{+\infty}x^{k}\left\{ B_{M}\frac{H_{M+1}(x)}{2x}\right\} \frac{e^{-x^{2}}}{\sqrt{\pi}}dx=(B_{M}/2)\intop_{-\infty}^{+\infty}x^{k-1}H_{M+1}(x)\frac{e^{-x^{2}}}{\sqrt{\pi}}dx=0\ \ if\ \ k-1<M+1$.
Thus, making $k-1\rightarrow k$ and $M+1\rightarrow M$ in the result
of appendix \ref{appx 2}, we then obtain the result.

\end{document}